\documentclass[fleqn,setspace,amsmath,amssymb]{article}
\usepackage{graphicx}
\usepackage{setspace}
\usepackage{amsmath}
\usepackage{amssymb}
\usepackage{rsfs}
 
\newcommand{\ra}{\rangle}
\newcommand{\pa}{\partial}
\newcommand{\la}{\langle}

\newcommand{\be}{\begin{equation}}
\newcommand{\ee}{\end{equation}}
\newcommand{\bea}{\begin{eqnarray}}
\newcommand{\eea}{\end{eqnarray}}

\begin{document}
\begin{titlepage}

\begin{flushright}
\today
\end{flushright}

\vspace{1in}

\begin{center}

{\bf Mechanism for spontaneous time reparametrization symmetry breaking in canonical gravity as origin of quantum energy measurement processes}

\vspace{1in}

\normalsize

{Eiji Konishi\footnote{E-mail address: konishi.eiji.27c@kyoto-u.jp}}

\normalsize
\vspace{.5in}

{\it Graduate School of Human and Environmental Studies, Kyoto University\\
 Kyoto 606-8501, Japan}
\end{center}

\vspace{1in}

\baselineskip=24pt
\begin{abstract}
We propose a mechanism for system-induced spontaneous time reparametrization symmetry breaking in canonical gravity.
We consider a model of extended canonical gravity, based on Sen's spinorial expression of the spatial-diffeomorphism-gauge independent part of the shift vector, with one additional long-range self-interacting massive spin-$1/2$ particle whose spin is coupled to Sen's spinor in the Hamiltonian.
This additional particle plays the role of creating the system-induced time reparametrization invariant potential.
We use this potential to break time reparametrization symmetry spontaneously.
We identify the symmetry breaking with the origin of the quantum energy measurement processes in the ensemble interpretation of quantum mechanics.
Our theory uses a novel interpretation of quantum mechanics with respect to quantum mechanical position uncertainties of particles.
The spatial-diffeomorphism-gauge independent part of the shift vector is attributed to this novel interpretation of quantum mechanics.

\end{abstract}

\vspace{.1in}


\vspace{.6in}


\end{titlepage}


\section{Introduction}

Since the advent of the Copenhagen interpretation of quantum mechanics\cite{Bohr,Bohr2}, it has been a great challenge to resolve the problems of measurement and the origin of the quantum state reduction that generates non-unitary time evolution of state vectors\cite{Neumann,Wigner,Everett,Decoherence,Decoherence2,Decoherence22,Decoherence3,Decoherence23,GRW,GRW2,Araki,Karolyhazy,Diosi,Diosi2,Penrose}.

In the ensemble interpretation of quantum mechanics\cite{dEspagnat}, there have been many earlier proposals for the physical realization of a {\it non-selective measurement} process (i.e., a decoherence process) that is a dynamical process and makes the quantum pure ensemble of a given state vector of a quantum system be equivalent to the classical mixed ensemble (i.e., the exclusive statistical mixture) of eigenstates of a discrete measured observable whose statistical weights in the ensemble of copies of the system are given by the Born rule\cite{Decoherence,Decoherence2,Decoherence22,Decoherence3,Decoherence23,GRW,GRW2,Araki,Diosi,Diosi2}.\footnote{The term {\it non-selective measurement} intuitively means that, after only this dynamical process, the events are in a classical mixed ensemble and a particular event is not yet selected from the events.}

However, no proposal has been given for the physical realization of the subsequent {\it event reading} process that is an informatical process and makes the classical mixed ensemble obtained by a non-selective measurement be the classical pure ensemble of an eigenstate of the measured observable.

Significantly, the distinction between the Copenhagen interpretation of quantum mechanics\cite{Bohr,Bohr2} and the decoherence program\cite{Decoherence23} is the physical realizability of this event reading process, that is, the final step of state reduction in the former.
For this reason, there has heretofore been no derivation of the Copenhagen interpretation as an entirely physical theory in the literature.

In this paper, we address a connection between these two measurement processes and time itself by extending {\it canonical gravity theory}, that is, the Hamiltonian formulation of classical general relativity, which is based on the Arnowitt--Deser--Misner (ADM) $3+1$ decomposition of space-time\cite{ADM}.
Specifically, we propose a possible mechanism for system-induced spontaneous time reparametrization symmetry (TRpS) breaking in canonical gravity theory and show that these two processes, with respect to energy measurements, accompany TRpS breaking as physical processes without modifying the Schr$\ddot{{\rm o}}$dinger equation.
In other words, we attempt to derive the Copenhagen interpretation as an entirely physical theory.
In the present investigation, we limit the focus to a mathematical model in which the mechanism works.

This paper is structured as follows.

In the next section, we extend canonical gravity by introducing one additional long-range self-interacting massive spin-$1/2$ particle.
In the Hamiltonian, the spin vector of this additional particle is coupled to the spatial-diffeomorphism-gauge independent part of the shift vector in the ADM $3+1$ decomposition of space-time under a novel interpretation of quantum mechanics with respect to quantum mechanical position uncertainties of particles.

In sections 3 and 4, we describe the mechanism of TRpS breaking and the two consequent measurement processes of desired quantum systems in our model.
In the mechanism of TRpS breaking, the additional particle plays the role of creating the system-induced time reparametrization (TRp) invariant potential, which is used to break TRpS spontaneously, and the result of Ref.\cite{KS} is essentially used to determine the ground state of the additional particle system.\footnote{For a comprehensive review of Ref.\cite{KS}, see Ref.\cite{KS2}.}

In the final section, we summarize our assumptions and overall results, discuss our model from three aspects, and discuss the novelty of our theory of measurement.

In Appendix A, we give a simple overview of the time concept in canonical gravity before its extension.

In Appendix B, we give a model of direct quantum measurements.

\section{The Model}

Canonical gravity theory is based on the ADM $3+1$ decomposition of a space-time $M$ having a metric tensor $g_{\mu\nu}$\cite{ADM}.
We choose the signature of $g_{\mu\nu}$ as $(+,-,-,-)$ and denote the time parameter and space-like hypersurfaces in their family by $t_0$ and $\Sigma_{t_0}$, respectively.

Here, $\Sigma_{t_0}$ has the induced metric tensor $q_{ab}(t_0,x^a)$ and the extrinsic curvature tensor $K_{ab}(t_0,x^a)$, which are precisely analogous to the position and the velocity of a particle, respectively.
In this paper, our interest lies in the spatial weak-field case
\begin{eqnarray}
(q_{ab}+\delta_{ab})^2&\approx&0\;,\label{eq:weak1}\\
K_{ab}^2&\approx &0\;.\label{eq:weak2}
\end{eqnarray}

In the ADM $3+1$ decomposition method, the square of the line element in $M$ is written as
\begin{equation}
ds^2=N^2c^2dt_0^2+q_{ab}(dx^a+N^a dt_0)(dx^b+N^bdt_0)\label{eq:ADM}
\end{equation}
for two actual variables: time-lapse function $N$ and shift vector $N^a$ (i.e., choice of space-time coordinate system).
For an overview of this method and intuitive interpretations of these actual variables $N$ and $N^a$, see Appendix A.

Here, we note two relations read from Eq.(\ref{eq:ADM}).
\begin{itemize}
\item The first relation is between the proper time $t$ and the clock time $t_0$
\begin{equation}
dt=Ndt_0\;.\label{eq:ele1}
\end{equation}
In our framework, we make the flow of clock time $t_0$ unique instead of making the flow of proper time unique.
This point will be discussed further in section 3.6.

\item The second relation is between the displacement of the spatial coordinate system $\delta x^a$ and the clock time $t_0$
\begin{equation}
\delta dx^a=N^a dt_0\;.\label{eq:ele2}
\end{equation}
From Eq.(\ref{eq:ele2}), the shift vector $N^a$ can be regarded as the {\it velocity} of the displacement of the spatial coordinate system $\delta x^a$.

\end{itemize}
These two relations are significant for later arguments.

\subsection{Sen's spinor}

Our model of extended canonical gravity is based on Sen's spinorial expression of a particular part of the shift vector.
The mathematical procedure for this expression is as follows.

We consider an $SL(2,C)$ two-component spinor variable $\lambda_{\alpha}$ on $M$, with three local degrees of freedom that appear in the real {\it null} four-vector $l_\mu\equiv \lambda_\alpha \bar{\lambda}_{\alpha^\prime}$\footnote{The {\it null four-vector} condition on a four-vector $l_\mu$ is $l^\mu l_\mu=0$.}, where $\bar{\lambda}_\alpha$ denotes the complex conjugate of $\lambda_\alpha$.\footnote{We lower and raise spinor indices by contraction with a symplectic form $i\sigma_2$.}
We restrict $\lambda_{\alpha}$ from $M$ to $\Sigma_{t_0}$ (here, $t_0$ is not specific).

We denote by $t^\mu$ the everywhere time-like future-directed unit four-vector field on $M$ which is normal to the family of $\Sigma_{t_0}$.
We restrict $t^\mu$ from $M$ to $\Sigma_{t_0}$ (here, $t_0$ is not specific) and give a Hermitian $SL(2,C)$ spinor representation, $t^{\alpha \alpha^\prime}$, for $t^\mu$ at each point in $\Sigma_{t_0}$ such that $t^{\alpha \alpha^\prime}\lambda_\alpha \bar{\lambda}_{\alpha^\prime}=t^\mu l_\mu$, in particular, holds.

Using $t^{\alpha \alpha^\prime}$, we give the Hermitian conjugate of $\lambda_{\alpha}$ by\cite{Sen2,Sen3}
\begin{equation}
\lambda^{\dagger\alpha}=\sqrt{2}t^{\alpha\alpha^\prime}\bar{\lambda}_{\alpha^\prime}\;.
\end{equation}
Here, note that the group that preserves the structure of a positive-definite Hermitian inner product $(\lambda,\eta)=\lambda^{\dagger \alpha}\eta_\alpha$ is $SU(2)$.
So, spinors $\lambda_\alpha$ defined over $\Sigma_{t_0}$ are in fact $SU(2)$ spinors on $\Sigma_{t_0}$.

Using this fact and following Sen\cite{Sen}, we express a particular part of the shift vector $N^a_{\rm phys}$ (introduced later) as a square of a dimensionless $SU(2)$ spinor $\lambda_\alpha$ on $\Sigma_{t_0}$:
\begin{eqnarray}
N^{(\alpha\beta)}_{\rm phys}=c\lambda^{\dagger (\alpha}\lambda^{\beta)}\;,\label{eq:NN}
\end{eqnarray}
where we use $()$ to indicate symmetric indices\cite{Review}.

Finally, we project the null vector $(\lambda^{\dagger\alpha}\lambda_\alpha/\sqrt{2},N^a_{\rm phys})$ to the three-vector $N^a_{\rm phys}$.
This projection of the null vector does not change $\lambda_\alpha$ because $\lambda_\alpha$ is an $SU(2)$ spinor on $\Sigma_{t_0}$.

Because this technique (\ref{eq:NN}) was first used in Ref.\cite{Sen}, we refer to this dimensionless $SU(2)$ spinor $\lambda_\alpha$ on $\Sigma_{t_0}$ as {\it Sen's spinor}.

\subsection{An additional particle}

Using the expression (\ref{eq:NN}), we introduce the spin-$1/2$ massive fermion field, called the {\it theta particle field}, on $\Sigma_{t_0}$ (here, $t_0$ is not specific) and extend canonical gravity.

We do this such that the two-dimensional spin Hilbert space, ${\cal V}$, of the theta particle matches the complex linear space of the Sen spinor $\lambda_\alpha$ in Eq.(\ref{eq:NN}).

In the present model, the theta particle is assumed to be a massive Dirac fermion and to be distinct from its own anti-particle.
In other words, the theta particle is assumed not to be truly neutral with respect to at least one kind of gauge potential.

In the present investigation, we hypothesize theta particles and anti-theta particles to be {\it cold dark matter} (CDM).
Each of these particles is assumed to have {\it no} direct interaction with the standard model elementary particles apart from its relevant gauge interactions and the origin of its mass.
We will elaborate on this hypothesis in the final section.

Taking advantage of the coldness of theta particles, we deal with only theta particles, which have velocities much lower than the speed of light.
Then, the theta particle field is approximately described by a non-relativistic Pauli $SU(2)$ two-component spinor field, $\Theta_\alpha$, on $\Sigma_{t_0}$ with dimensions [L$^{-3/2}$].

\subsection{The Hamiltonian}

In our model, we consider the gravity sector, the theta particle sector, and the matter sector, $\Psi$, for low-energy phenomena.
Namely, our model consists of three sectors.

In the ADM $3+1$ decomposition of space-time, the total Hamiltonian consists of four parts:
\begin{equation}
H_{\rm tot}=H_{\rm grav}+H_{\rm spin}+H_c+H_{\rm mat}\;,\label{eq:ModelH}
\end{equation}
where
\begin{enumerate}

\item[P1] $H_{\rm grav}$ is the gravitational Hamiltonian.

\item[P2] $H_{\rm spin}$ is the Hamiltonian of the theta particle field itself as that of a spin model.

\item[P3] $H_c$ is the interaction Hamiltonian between the theta spin and the Sen spin.

This interaction Hamiltonian is also the TRp invariant potential.
Its density is not reducible to a form written in terms of canonical variables.

\item[P4] $H_{\rm mat}$ is the Hamiltonian of the matter sector $\Psi$.

\end{enumerate}

We define the part $H_{\rm grav}$ in Appendix A.
In the rest of this subsection, we define our original parts, namely, $H_{\rm spin}$ and $H_c$.

\subsubsection{The part $H_{\rm spin}$}

We define the spin density of the theta particles by
\begin{equation}
\tau^{(\alpha\beta)}=\Theta^{\dagger(\alpha}\Theta^{\beta)}
\end{equation}
in the $SO(3)$ vector spin model, in which all variables are treated as their field amounts.
Incorporating this variable, we model the Hamiltonian of the theta particle field $\Theta_\alpha$ itself (excluding the gravitational interaction part) as that of a spin model with a spin-spin interaction that is {\it long-range} (like the gravitational and Coulomb interactions) and {\it ferromagnetic} (attractive for two aligned spins and repulsive for two opposite spins) via a massless wave of a continuous succession of spin-spin interactions, which propagates in the {\it additional} (dimensionless) Sen spinor $\Lambda_\alpha$ that belongs to ${\cal V}$,
\begin{equation}
H_{{\rm{spin}}}=\int_{\Sigma_{t_0}}(Nh_{{\rm{spin}}}-N_a h^a_{{\rm spin}})dS\;,\label{eq:Hspin}
\end{equation}
where $dS$ is the volume element defined by $dS_\mu=t_\mu dS$.
Here, in the Newtonian limit,
\begin{eqnarray}
{{h}}_{{\rm{spin}}}&\approx&\frac{i\hbar}{2m}\pa_a
{{\pi}}_\alpha\pa^a\Theta_\beta\delta^{\alpha\beta}-J\int_{\Sigma_{t_0}}\frac{1}{|x-x^\prime|}{{\tau}}_a(x){{\tau}}^a (x^\prime)dS^\prime\;,\label{eq:XY}\\
h^a_{{\rm spin}}&\approx&-\pi_\alpha \pa^a \Theta_\beta\delta^{\alpha\beta}
\end{eqnarray}
for a theta particle field ${\Theta}_\alpha$ with mass $m$, canonical momentum ${\pi}_\alpha=i\hbar\bar{\Theta}_\alpha$, and a positive-valued long-range decay factor $-2J/|x-x^\prime|$, with a constant $-J>0$, that generates an inverse square law force.

To derive the interaction part of Eq.(\ref{eq:XY}), we introduce a dimensionless vector field
\begin{equation}
M^{(\alpha\beta)}=\Lambda^{\dagger (\alpha}\Lambda^{\beta)}
\end{equation}
for the additional Sen spinor $\Lambda_\alpha$.
In the Newtonian limit, we assume that $M^a$ is an {\it auxiliary field} (i.e., without its time-derivative terms in the Lagrangian density) governed by the Lagrangian density of spin-spin interactions\footnote{For the second term in Eq.(\ref{eq:XY1}), refer to footnote 6b in Ref.\cite{HK}.}
\begin{equation}
l_\Lambda\approx 2J_0\tau_aM^a-A\pa_a M_b\pa^a M^b\;,\label{eq:XY1}
\end{equation}
where $J_0$ and $A>0$ are constants.
We can derive the interaction part of Eq.(\ref{eq:XY}) from Eq.(\ref{eq:XY1}).
By doing so, we also obtain the relation $-J=J_0^2/(4\pi A)$.

\subsubsection{The part $H_c$}

Next, in our model, the theta spin interacts with the Sen spin $\lambda_\alpha$ by a strong physical coupling, $h_c$, times $N$ in their own variable spaces:
\begin{equation}
H_c\approx -2J_c\sum_{\psi_0} w_{\psi_0}\int_{\Sigma_{t_0}}\tau_aN_{\rm phys}^a|_{\psi_0} dS\;,\label{eq:Hc}
\end{equation}
where $-J_c>0$ is a constant, and $H_c=\int_{\Sigma_{t_0}}Nh_cdS$.
Here, the density $h_c$ of the interaction Hamiltonian $H_c$ is not reducible to a form written in terms of canonical variables.

In $H_c$, the term $N_{\rm phys}^a$ is the spatial-diffeomorphism-gauge independent part (i.e., the {\it physical} part) of the shift vector
\begin{equation}
N^a=N^a_{\rm phys}+N^a_{\rm rest}\label{eq:Nphys}
\end{equation}
and $N_{\rm phys}^a|_{\psi_0}$ is defined for each {\it reduced particle pure state} $\psi_0$ with statistical weight $w_{\psi_0}$ in the mixed state of each particle obtained by the partial trace of the rest particles in the whole system of all particles.

Here, the {\it velocity} $N^a_{\rm phys}$ gives a physical spatial displacement, within $M$, of each $\psi_0$ and $\Sigma_{t_0}$, accompanying the time elapse via its couplings to supermomenta, $\{h^a\}$, in the Hamiltonians.
In contrast, $N^a_{\rm rest}\equiv -N^a_{\rm phys}+N^a$ is unphysical and depends on the choice of a gauge $N^a_{t_0}$ that is independent of any $\psi_0$.

In short, whereas gauge $N^a$ represents an arbitrary choice of the spatial coordinate system, $N^a_{\rm phys}$ represents a physical displacement of the spatial coordinate system.

We will define $N_{\rm phys}^a|_{\psi_0}$ in Eq.(\ref{eq:Hc}) by Eq.(\ref{eq:DPO1}) in section 3.2 after we present our novel interpretation of quantum mechanics with respect to quantum mechanical position uncertainties in section 2.4.

Here, we observe two points from this definition of $H_c$:
\begin{itemize}

\item $H_c$ is invariant under the TRp $t_0\to t_0^\prime=F(t_0)$ at the level of canonical equations because the combination $H_cdt_0$ is invariant under TRp.

\item 

A spin-spin interaction is between two spins with a common two-dimensional spin eigenspace.
From this fact, the anti-theta spin does not have the coupling $H_c$.

\end{itemize}

This strong ferromagnetic coupling $H_c$ plays a definite role in our model: if $-2J_c|N_{\rm phys}^a|$ is significant, $H_c$ forces the theta spin to align in the direction of $N_{\rm phys}^a$ (to decrease the value of energy $H_c$) in the dissipative and adiabatic (due to the low velocity) motion of the theta particle.
This is because $H_c$ does not change the direction and vector modulus of the Sen spin.
If there is no dissipation of energy, the theta spin precesses around the axis $N_{\rm phys}^a$.

\subsubsection{On the space-time diffeomorphism invariance}

Finally, we must examine the space-time diffeomorphism invariance of the total Hamiltonian $H_{\rm tot}$.
There are two points to this with respect to $H_c$ and $H_{\rm spin}$.

First, $H_c$ violates the super-Hamiltonian and supermomentum constraints due to the non-canonical nature of the density $h_c$ of $H_c$; this violation is just a {\it quantum coherence effect induced on the theta particle sector} by $\psi_0$ (we will derive this {\it quantum coherence effect} in section 3.2) and has not yet been detected.

For this violation, in this model, the space-time diffeomorphism invariance of the total Hamiltonian $H_{\rm tot}$ is partially and explicitly broken.
Namely, in this model, the final result of time evolution, of the data that include the canonical variables of $\Sigma_{t_0}$, depends on the choice of a path of $\Sigma_{t_0}$ in $M$, that is, the choice of a diffeomorphism gauge of a part of space-time $(N_{t_0},N_{t_0}^a)$ under fixing of both ends of paths and the data at the initial time.\footnote{This consequence is a known result.
See Ref.\cite{Geometry}.}

However, TRpS of the combination $H_{\rm tot}dt_0$ (namely, TRpS of the canonical time evolution) is not explicitly broken.
In other words, for a specifically chosen path of $\Sigma_{t_0}$ in $M$, the time evolution of the data along it does not depend on the choice of a clock $t_0$ (see also section 3.6).

Second, in the form presented in Eq.(\ref{eq:Hspin}), $H_{{\rm spin}}$ manifests a global $SU(2)$ spin rotation symmetry.

This global symmetry gives rise to the global $SO(3)$ spatial rotation symmetry and is gauged by the additional Sen spinor $\Lambda_\alpha$ that does not explicitly appear in $h_{{\rm spin}}$ by substituting the covariant derivative for the derivative in Eq.(\ref{eq:XY1}).

This gauging makes $H_{\rm spin}+H_c$ and thus $H_{\rm tot}$ be spatial-diffeomorphism invariant with respect to intrinsic coordinates for every slice $\Sigma_{t_0}$ of $M$.

Since spatial coordinates are not canonical variables, the choice of a diffeomorphism gauge of the whole of space-time $(N_{t_0},N^a_{t_0})$ is arbitrary (i.e., not restricted by the dynamics).

\subsection{A novel interpretation of quantum mechanics}

Our framework is based on the following novel interpretation of quantum mechanics with respect to quantum mechanical position uncertainties.

\begin{figure}[htbp]
\begin{center}
\includegraphics[width=0.48 \hsize,bb=5 5 259 258]{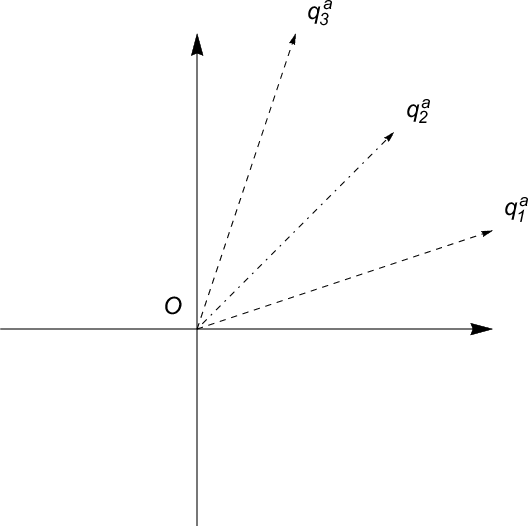}
\includegraphics[width=0.48 \hsize,bb=5 5 258 259]{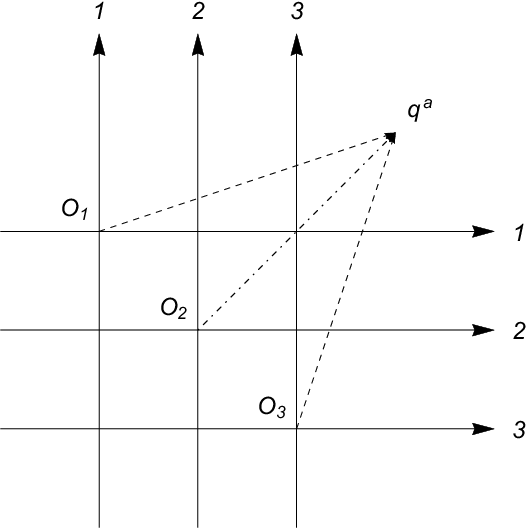}
\end{center}
\caption{Planar slice.
The left panel illustrates the conventional interpretation of one particle quantum mechanical position uncertainty.
In the left panel, the spatial coordinate system with origin $O$ is definite, whereas the uncertainty is owned by the position $q^a$ of the particle, which has three varieties ($q_1^a$, $q_2^a$, and $q_3^a$) in the figure.
In contrast, the right panel illustrates the novel interpretation of one particle quantum mechanical position uncertainty.
In the right panel, the position $q^a$ of the particle is definite, whereas the uncertainty is owned by the coordinate values $\xi^a=q^a$ measured in three distinct spatial coordinate systems with origins $O_1$, $O_2$, and $O_3$ corresponding to, respectively, positions $q_1^a$, $q_2^a$, and $q_3^a$ of the particle in the left panel.
When three coordinate systems with the particle are equally weighted by a complex number (whose absolute square is the existence probability) in superposition, the coordinate system with the origin $O_2$ is the average of the coordinate systems with the origins $O_1$, $O_2$, and $O_3$.}
\end{figure}

In this interpretation, the positions, $\{q^a\}$, of the localizable $n$ {\it particles} in a quantum matter system $\Psi$\cite{NW,NW2,BB} are, originally, definite (i.e., with no uncertainty) variables {\it in the $3n$-dimensional spatial configuration space} of these particles.\footnote{We treat a composite particle with the center of mass position as {\it a particle in our sense} if its elementary particles cannot be excited.}

Instead of having uncertain original positions, the uncertainties of $\{q^a\}$ at time $t_0$ are owned by $n$ space-like hypersurfaces $\{\Sigma_{t_0}\}$ related to the spatial configuration space of $\Psi$ as the uncertainties of the variables $\{\xi^a\}$ at the points $\{\xi^a=q^a\}$ occupied by the $n$ particles (see Fig.1, which is written for $n=1$).

Here, $\xi^a=\xi^a_{t_0}$ is a coordinate system in $\Sigma_{t_0}$, and its average (that is, its quantum mechanical expectation value, precisely defined in Eq.(\ref{eq:Delta})) is assumed to match the spatial coordinate system obtained by the ADM $3+1$ decomposition of space-time:
\begin{equation}
\la \xi^a_{t_0}\ra=x^a+\int^{t_0}_{t_{0,{\rm ini}}} N^a_{t_0^\prime}(x)dt_0^\prime\;.\label{eq:xiafin}
\end{equation}

Next, the equation $\xi^a=q^a$ (or the symbol $\xi^a|_{q^a}$) refers to the three coordinate values at the geometric point $q^a$ in $\Sigma_{t_0}$, measured in the coordinate system $\xi^a$.

[The uncertainties of $\xi^a$ are {\it physical} at the point $\xi^a=q^a$ occupied by a particle because of this interpretation and the fact that $q^a$ are canonical variables.
In contrast, the uncertainties of $\xi^a$ are {\it unphysical} at the other points $\xi^a\neq q^a$ occupied by no particle because $\xi^a$ are not canonical variables.]

In this interpretation, a single-particle position superposition of the spatial coordinate systems is logically parallelly generalized to a non-factorizable multi-particle position superposition of the spatial configuration spaces.
This interpretation requires the first-quantized treatment (in the Fock space of direct sums of zero/single/multi-particle pure states) of identical particles whose positions are originally definite in their spatial configuration space.

We emphasize two points.
First, as can be seen from Fig.1, the wave function
\begin{equation}
\psi(\xi^a=q^a,t)=\la \xi^a=q^a|\psi(t)\ra\label{eq:wavef}
\end{equation}
of a particle with a state vector $|\psi(t)\ra$ (in the Schr$\ddot{{\rm{o}}}$dinger picture) in this interpretation is of course the same as that in the conventional interpretation.
Here, $\{|\xi^a=q^a\ra\}$ are the simultaneous eigenstates of the three canonical position operators $\widehat{\xi^a|_{q^a}}$ of this particle, and $\widehat{\xi^a|_{q^a}}$ are defined by the procedure of canonical quantization.
Second, the relevant difference between these two intepretations is the origin of the quantum mechanical position uncertainties.

The purpose of adopting this novel interpretation is to convert the quantum mechanical position uncertainties of particles in a desired quantum matter system into the overall aligned distribution of the Sen spin (see section 3.2).
In the conventional interpretation, such a conversion never happens, and then the TRp invariant potential $H_c$ is trivially zero.
This novel interpretation is at the heart of the symmetry breaking mechanism.

\section{Symmetry Breaking}
In this section, we explain the symmetry breaking mechanism.

\subsection{Main statement}

This paper's main statement is that, in our model, {\it TRpS can be spontaneously broken in time-dependent processes of a weakly coupled (i.e., almost not entangled) and spatially macroscopically coherent quantum matter system $\Psi$} (a {\it quantum mechanically macroscopic length} is typically above the micrometer scale) {\it and this TRpS breaking leads to energy measurement processes of $\Psi$.}

Throughout the discussion of symmetry breaking, we use the term {\it coherence} with the meaning of the coherent property of superposed de Broglie waves of matter and superposed photons (note Refs.\cite{NW,NW2,BB}) where the particle number and the phase are virtually definite.\footnote{Coherent pure states of photons can be localized without limitation.
The fully localized photon number density can be obtained for these pure states.
For details, see Ref.\cite{BB}.}

In a coherent quantum pure state realized over the spatial domain, $V$, where the system $\Psi$ is bound, there is an overwhelming majority of bosons with a specific mode $k_0$.
Then, due to the position-momentum uncertainty relation, in this pure state, the three position uncertainties of $\Psi$ {\it in itself} are realized with the dimensions of this cut-off spatial domain $V$ of the plane wave with the mode $k_0$.

\subsection{Initial state of the theta particle system in a coherence domain}

Now, we assume a weakly coupled quantum matter system $\Psi$ of $n$ identical particles (bosons) in a pure state $|\Psi\ra$, with the coherence of $\Psi$ spherically limited by a macroscopic radius $\ell$ at a time $t_0$.\footnote{In this, $n$ has a fluctuation $\Delta n$ around the average $n_0$ such that $n_0\gg \Delta n \gg 1$ holds.
For a Bose-Einstein condensate, to give a prominent example, $\Delta n\sim \sqrt{n_0}$ holds.}

At the same time, {\it we project the $3n$-dimensional spatial configuration space of these particles onto the three-dimensional space $\Sigma_{t_0}$}.
For a multi-particle pure state, we {\it trace out} the rest particles of the representative particle before this projection; for each permutation-symmetric whole pure state of $\Psi$ with a definite particle number $n$, this partial trace gives rise to a mixed state of the representative particle that consists of ${\mathscr{N}}_n$ reduced pure states.

In this setting, we will show the existence of the macroscopically, in the quantum mechanical sense,
overall aligned
distribution of total Sen spins over the time-dependent {\it coherence domain} $V$ of $\Psi$ in $\Sigma_{t_0}$ at time $t_0$ (see Eq.(\ref{eq:PhysSen})).

The two key specific assumptions (that is, the setup) for showing this existence concern the system $\Psi$.

First, the order parameter of the {\it coherence} of the system $\Psi$ over the coherence domain $V$ is assumed to take a non-zero value such as
\begin{equation}
n^{-1}_0\int_V|\la \widehat{\Psi}(x)\ra|^2dS\sim 1\;.\label{eq:COP}
\end{equation}
So, the system $\Psi$ {\it in itself} can be well described by the {\it macroscopic wave function} $\la \widehat{\Psi}(x)\ra$.

Second, the radius $\ell$ of the coherence domain $V$ is assumed to be close to the physical uncertainties of the coordinate values $\Delta \xi^a=\Delta \xi^a_{t_0}$ (defined for the representative particle in a pure state) at the originally definite position of the representative particle, that is, a point $\xi^a=q^a$.
This point $\xi^a=q^a$ is identified between the coordinate systems $\{\xi^a\}$ under superposition.
Specifically, these physical uncertainties $\Delta \xi^a$ are taken to be the root mean square error of the $q^a$-values in the coordinate systems $\{\xi^a\}$ from the $q^a$-value in the {\it average coordinate system} $\la \xi^a\ra$ at $\xi^a=q^a$:
\begin{equation}
\Delta \xi^a\equiv\sqrt{\la (\xi^a-\la \xi^a\ra)^2\ra}\ \ {\rm at}\ \ \xi^a=q^a \label{eq:Delta}
\end{equation}
in this identification of the point $\xi^a=q^a$.
In Eq.(\ref{eq:Delta}), averages are taken by using the wave function (\ref{eq:wavef}) of each pure state in the density matrix related to the representative particle $\widehat{\varrho}_{\rm rep}\equiv{\rm tr}_{\rm rest}|\Psi\ra\la \Psi|$.

Now, we identify the superposed coordinate systems $\{\xi^a\}$ (refer to Fig.1).
Then, the point $\xi^a=q^a$ is distributed over $V$ under superposition, and we obtain that
\begin{equation}
n^a\ell\chi_V \approx{{{N}_{\rm phys}^a}\Delta t_0}\label{eq:DPO1}
\end{equation}
and
\begin{equation}
\overline{n^a}\ell\chi_V \approx{\overline{{N}_{\rm phys}^a}\Delta t_0}\label{eq:DPO}
\end{equation}
for the $w_{\psi_0}$-weighted arithmetic mean $\overline{X}$ of $X$ over all $\sum_n{\mathscr{N}}_n$ reduced particle pure states, a unit vector $n^a_{t_0}$ (here, $|\overline{n^a}|\sim 1$ due to Eq.(\ref{eq:COP})), the characteristic function $\chi_V=\chi_V(x)$ of the coherence domain $V$, $N^a_{\rm phys}=N^a_{{\rm phys},t_0}(x)$, and the TRp invariant non-mechanical Mandelstam-Tamm time uncertainty, $\sigma$, of ${\xi^a}$\cite{Mandelstam,Messiah}:
\begin{eqnarray}
\sigma&=&N\frac{|\Delta \xi^a|}{|N^a_{\rm phys}|}\label{eq:SIGMA}\\
&\overset{\Delta t_0}{\equiv}&N\Delta t_0\;.
\end{eqnarray}
Here, $\sigma$ is constant over $V$ and is the characteristic time for the rate of the {\it physical part of} change of the coordinate system $\la\xi^a\ra$ in $V$ (see Eq.(\ref{eq:Nphys})) under superposition at the time $t_0$\cite{Messiah}.
As a result, we obtain
\begin{eqnarray}
|\overline{{N}_{\rm phys}^a}|&\approx&\frac{|\overline{n^a}|\ell \chi_V}{\sigma}N\label{eq:PhysSen}\\
&\overset{\kappa}{\equiv}&{\kappa}\chi_V N\;.\label{eq:kappadef}
\end{eqnarray}

\begin{figure}[htbp]
\begin{center}
\includegraphics[width=0.5 \hsize,bb=5 5 258 259]{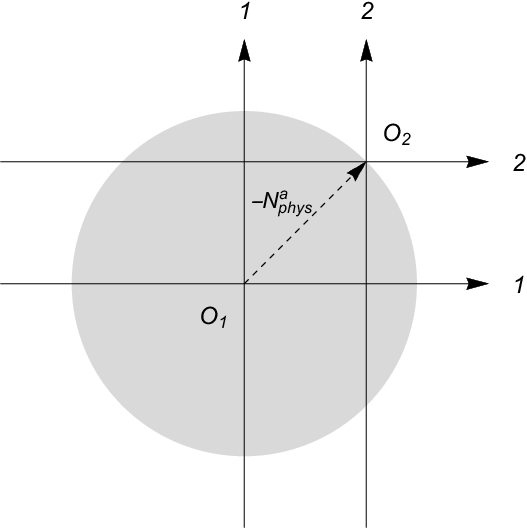}
\end{center}
\caption{Planar slice.
In this figure, the origin of the average spatial coordinate system defined for a reduced particle pure state is physically displaced from $O_1$ to $O_2$ through the coherence domain $V$ (the gray domain).
Here, we set $\la \xi^a\ra=0$ at $\xi^a=q^a$ at the instance when the origin is at $O_1$.
This displacement has {\it velocity} $-N_{\rm phys}^a$ and is done within the Mandelstam-Tamm time uncertainty $\sigma$.}
\end{figure}

In Eq.(\ref{eq:DPO1}), $n^a$ is the global direction of the {\it group velocity} $N_{\rm phys}^a$ (defined over $V$) of the sets of three coordinate values $\{\xi^a\}$ at the originally definite position of a particle (that is, a point $\xi^a=q^a$) measured in the respective identified coordinate systems $\{\xi^a\}$ under superposition with the uncertainties $\Delta \xi^a$ (see Fig.2).

(Note that whereas $N^a$ must give a smooth spatial displacement that is well defined over $\Sigma_{t_0}$ to maintain the well-definedness of the canonical equations, its parts $N^a_{\rm phys}$ and $N^a_{\rm rest}$ need not do so by themselves.)

Here, $n^a$ is randomly selected with mean zero due to the $SU(2)$ spin rotation symmetry of the Sen spin system, and needs to be definite for each {\it coherent segment of $\Psi$} (i.e., each subsystem pure {\it state} of a large number, fluctuating in superposition, of bosons with a specific mode and a specific phase) within the coherence domain $V$.
This is because such a segment can, in itself, be described by a macroscopic wave function of a single position variable.
In contrast, for each incoherent segment of $\Psi$ within the domain $V$, the $w_{\psi_0}$-weighted arithmetic mean of $n^a$ over the segment vanishes due to the randomness in each selection of $n^a$.

In this derivation of $N_{\rm phys}^a$ from our novel interpretation of quantum mechanics, we note three distinct points.
\begin{itemize}
\item The time-energy uncertainty relation is not applied to the Mandelstam-Tamm time uncertainty $\sigma$.

This is because $\sigma$ is not defined through Hamiltonian mechanics (note that $\xi^a$ are not canonical variables) but is the {\it statistical dispersion of equivalent proper time increment trials} given at a time $t_0$ (see also section 3.6).

\item When two or more quantum matter systems $\{\Psi\}$ of different kinds of particles coexist in $V$, each distinct quantum matter system is assumed to contribute independently to the coupling $H_c$.
Specifically, the integrand of $H_c$ can be written as $-2J_c\sum_{\Psi}\tau_aN_{\rm phys,\Psi}^a$.

This assumption further applies to identical particles in each $\Psi$ at the level of weighted and reduced particle pure states (see Eq.(\ref{eq:Hc})).

\item The direction of $\overline{n^a}$ in Eq.(\ref{eq:DPO}) is definite while Eq.(\ref{eq:COP}) is satisfied.

This is because $N_{\rm phys}^a$ is an auxiliary field in our definition, and in the coupling $H_c$, to rotate $N_{\rm phys}^a$ by itself means changing energy without cause.

\end{itemize}

\begin{figure}[htbp]
\begin{center}
\includegraphics[width=0.5 \hsize,bb=5 5 255 151]{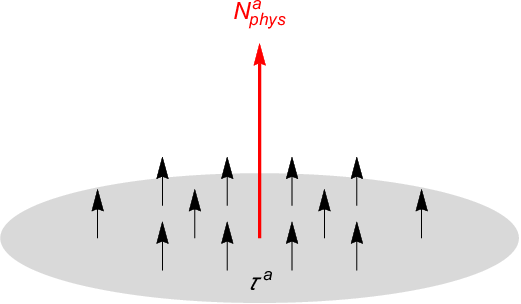}
\end{center}
\caption{
Planar slice.
In a coherence domain $V$ (the gray domain), the distribution of the Sen spins $N_{\rm phys}^a$ (summed up over all of the reduced particle pure states and represented by the long thick red arrow) is overall aligned.
Due to the ferromagnetic coupling $H_c$ between the Sen spins and theta spins, the distribution $\tau^a$ of the theta spins (the short, thin black arrows) in $V$ aligns in the direction of $N_{\rm phys}^a$ as a result of the dissipation of energy of the theta spin system.
Then, due to the attractive and long-range self-interactions $H_{\rm spin}$ between these aligned theta spins, the collisionless equilibrium state of the theta spin system will be realized within $V$ (see section 3.3).}
\end{figure}

Now, we define the {\it magnetization} of a system of spins by {\it the vector modulus of arithmetic mean of the spin vector over this system}.
Then, the configuration of Sen spins in Eq.(\ref{eq:DPO}) generates a magnetized distribution $\tau^a$ of spin vectors of theta particles via the ferromagnetic spin-spin coupling $h_c$ that tends to align $\tau^a$ in the direction of $N_{\rm phys}^a$ (see Fig.3).

As an illustration, if we take the strong-$J_c$ limit of Eq.(\ref{eq:Hc}) in the model setting, in which $h_c$ is the dominant energy of the theta particle system within the coherence domain $V$, $\tau^a$ of essentially all of the theta spins will relax to align in the direction of $N_{\rm phys}^a$.
Such relaxation is the result of interactions between each theta particle and its environmental particles, that is, a dissipation effect.

Here, an important point is that, due to Eq.(\ref{eq:XY}), interactions between these aligned theta spins in the magnetized system are {\it attractive}.

\subsection{Ground state of the theta particle system}

Once such a locally isolated theta particle system with aligned spins accompanies a spatially macroscopically coherent quantum matter system $\Psi$, the theta particles in it settle down to their local {\it collisionless equilibrium} state as a result of the so-called {\it phase mixing}\footnote{Here, {\it phase mixing} means time-dependent coarse-graining of the $\mu$-space distribution of a long-range interacting system facilitated by strong oscillation of the self-consistent mean-field potential\cite{KS2,LB}.} before they approach their local Boltzmann-Gibbs equilibrium state (i.e., their local {\it collisional equilibrium} state).
This is due to the {\it long-range nature of the attractive interactions} in $h_{{\rm spin}}$\cite{KS,KS2,LB,PL1,PL12,Book,PR1}.

We infer the following consequence based on recent results in the mean-field analysis of the three-dimensional self-gravitating model\cite{3D} and the core-halo structure of the collisionless equilibrium state in the simplest benchmark model for long-range interacting systems of the type given in Eq.(\ref{eq:XY})\cite{KS,KS2,PL1,PL12} and the universality of Lynden-Bell statistics for the collisionless equilibrium in long-range interacting systems\cite{LB}.
That is, in general, the collisionless equilibrium one-particle distribution, $f(\varepsilon)$, of the one-particle energy $\varepsilon$\footnote{Here, the one-particle energy is a function $\varepsilon=\varepsilon(q^a,p^a)$ of the one-particle position $q^a$ and momentum $p^a$ including the self-consistent mean-field potential of this system.}, in the mean-field treatment, does not undergo a sufficiently violent collisionless relaxation process (i.e., sufficiently violent phase mixing) to reach the single Lynden-Bell ergodic equilibrium but is realized as a less ergodic {\it superposition} of two independent, low-energy dense {\it core} and high-energy {\it halo}, Lynden-Bell coarse-grained distributions
\begin{equation}
f^{(i)}(\varepsilon)=\frac{\eta^{(i)}}{\exp(\beta^{(i)}(\varepsilon-\mu^{(i)}))+1}\;,\ \ i=1,2\;.\label{eq:DLB}
\end{equation}
Here, $\eta$ is the diluted $\mu$-space density\cite{KS,KS2}, and two Lagrange multipliers appear: inverse temperature $\beta$ for the total energy and chemical potential $\mu$ for the total mass conservation.

This collisionless equilibrium distribution was discovered and called the {\it double Lynden-Bell distribution} in Ref.\cite{KS,KS2}, and has, in particular, two {\it coexisting} chemical potentials of one kind of particle in the superposition.
(The masses that have completely escaped from the main cluster are excluded from the local system.)
Here, due to the Pauli exclusion principle for theta particles (i.e., theta fermions) and the alignment of the initial theta spins, we consider an initial $\mu$-space distribution with a single non-zero fine-grained level $\eta_0=1/h^3$, where $h$ is the Planck constant.

\begin{figure}[htbp]
\begin{center}
\includegraphics[width=0.5 \hsize,bb=4 5 258 259]{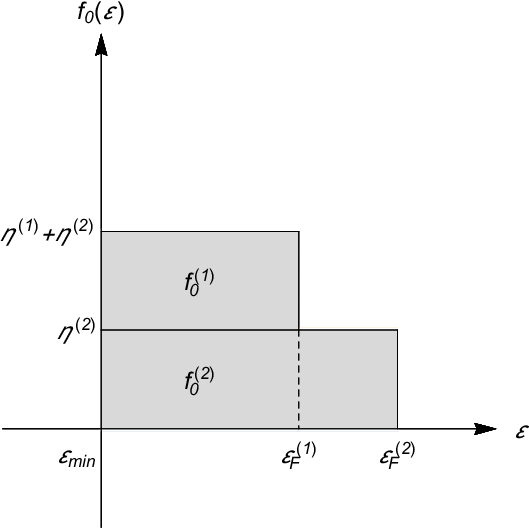}
\end{center}
\caption{Double Lynden-Bell distribution of the ground state (\ref{eq:GSdis}) of the theta particle system.
This distribution is a two-step functional of the one-theta-particle energy $\varepsilon$.}
\end{figure}

The ground state one-particle distribution of the theta particle system is a {\it two-step functional} of the one-particle energy
\begin{equation}
f_0(\varepsilon)=\sum_{i=1,2}f_0^{(i)}(\varepsilon)=\sum_{i=1,2}\eta^{(i)}\theta(\varepsilon_F^{(i)}-\varepsilon)\label{eq:GSdis}
\end{equation}
for the Heaviside step functional $\theta(\varepsilon)$ and the {\it Fermi energy} $\varepsilon_F\equiv \mu|_{\beta \to \infty}$ of each Lynden-Bell distribution at zero temperature (see Fig.4).

Here, we make two remarks about the {\it collisionless equilibrium}.

First, in the collisionless equilibrium of a long-range interacting system, as a result of phase mixing, the variable of the coarse-grained distribution is the one-particle energy $\varepsilon$ only: $f=f(\varepsilon)$\cite{Book}.

The second remark concerns the persistence of this equilibrium.
A general property of a long-range interacting system means that, in the thermodynamic limit (i.e., the limit that the particle number $\nu$ tends to infinity while fixing the energy per particle $E/\nu$), the lifetime of the collisionless equilibrium, $t_c$, diverges.\footnote{In the system of aligned theta spins, $t_c\sim(\nu/8\ln \nu)t_f$ holds for the free-fall time $t_f\sim (-J\nu/mV)^{-1/2}$\cite{Galaxy}.}
In the exact thermodynamic limit, this system never approaches the Boltzmann-Gibbs equilibrium from the collisionless equilibrium\cite{Book,PR1,BH}.

\subsection{Ground state of the total system}

In the ground state $f_0$ of the theta particle system, $H_c$ can be written as
\begin{eqnarray}
V_c&=&-2J_c\sum_{\psi_0}w_{\psi_0}\int_{\Sigma_{t_0}}\tau_a(x)N^a_{{\rm phys},t_0}|_{\psi_0}(x) dS \\
&\approx&2J_c{n_0}\kappa\sum_{i=1,2}\nu \int_V\rho^{(i)}_0(x) N_{t_0}(x)dS\;,\label{eq:pot}
\end{eqnarray}
where $\kappa$ defined by Eq.(\ref{eq:kappadef}) is TRp invariant, $\nu$ is the particle number of the theta particle system, and $\rho^{(i)}_0$ is the position distribution of the coarse-grained $\mu$-space distribution $f^{(i)}_0$.
$\rho^{(i)}_0$ is obtained by integrating out the momentum variables of $f^{(i)}_0$.
The form of $f^{(i)}_0$ is determined by the {\it Fermi energy} $\varepsilon_F^{(i)}$ and the form of the self-consistent mean-field potential of the second term of Eq.(\ref{eq:XY}).

Here, $V_c$ is the TRp invariant potential of the total system at the level of canonical equations, and we use this potential to break TRpS spontaneously (i.e., we show that the ground state of this TRp invariant potential is not TRp invariant).
The main property of this potential is that $V_cdt_0$ gives two distinct invariance classes $(t_0,\rho^{(i)}_0N)$ ($i=1,2$), with respect to TRp, for two different time lapse densities $\rho^{(i)}_0N$ in the coarse-grained integrand of $V_c$, respectively.
Owing to their differing TRp invariance classes, these two time lapse densities $\rho^{(i)}_0N$ define different proper times $t^{(i)}$ of $f^{(i)}_0$ and corresponding time lapses $N^{(i)}$.

This property of $V_c$ is confirmed by noting that, in the particle description (i.e., in the fine-grained description), $V_c$ is written as
\begin{eqnarray}
V_c\approx 2J_c n_0\kappa\sum_{i=1,2}\sum_{q^{(i)}\in \nu \rho_0^{(i)}} N_{t_0}(q^{(i)})\;,\label{eq:pd}
\end{eqnarray}
where time lapses $N_{t_0}(q^{(i)})$ correspond to the time lapse densities $\rho^{(i)}_0N$ in Eq.(\ref{eq:pot}).
Here, we emphasize that this coarse-graining is applied not to the proper time $t$ but to the integrand of $V_c$.

Due to this main property of $V_c$, the ground state one-particle distribution function of the total system is in the form
\begin{equation}
{\boldsymbol f}_0((q,N_q),p)=\sum_{i=1,2}{\boldsymbol f}_0^{(i)}((q,N_q),p)\;.\label{eq:Gstate}
\end{equation}
This expression means that for $i=1,2$,
\begin{eqnarray}
f_0^{(i)}(q,p)&=&\la {\boldsymbol f}_0^{(i)}((q,N_q),p)\ra_{N_q}\;,\label{eq:disteq}\\
\rho_0N^{(i)}&{=}&({\rho^{(i)}_0}/{r^{(i)}})N{=}\rho_0{\la N\ra_0^{(i)}}\;,\label{eq:ti}
\end{eqnarray}
where $f^{(i)}_0(q,p)$ is the $i$-th $\mu$-space Lynden-Bell distribution function of theta particles with the fixed (i.e., aligned) spin part at zero temperature (see Eq.(\ref{eq:GSdis})), $r^{(i)}$ is the fraction of theta particles in $\nu f_0^{(i)}$ relative to $\nu$, and $\la\cdot\ra_{X}$ and $\la \cdot\ra_0^{(i)}$ represent the integrating out of $X$ and the average using $\phi^{(i)}_0(N_q)=\la {\boldsymbol f}_0^{(i)}((q,N_q),p)\ra_p/\rho_0^{(i)}(q)$ as the {\it functional} weight, respectively.

In Eq.(\ref{eq:disteq}), we apply the definition of a {\it distribution function} in statistical mechanics as the weight in the averages of variables to the theta particle system.
In Eq.(\ref{eq:ti}), for $i=1,2$, the first and the second equalities for one theta particle define, respectively, $N^{(i)}$ and the $N_q$-dependence of ${\boldsymbol f}_0^{(i)}$; $\rho_0N^{(i)}$ rewrites the coupling $(\rho_0^{(i)}/r^{(i)})N$ per theta particle in $\nu\rho_0^{(i)}$ to that in the theta particle system $\nu\rho_0$.

\subsection{Symmetry breaking: the mainstay argument based on Eq.({\ref{eq:GSdis}})}

Now, we show that though $V_cdt_0$ is TRp invariant, the ground state ${\boldsymbol f}_0$ is not TRp invariant due to the inconsistency of TRp invariance with the theta particle part of this ground state: non-zero $N^{(1)}-N^{(2)}$ obtained from Eq.(\ref{eq:GSdis}) breaks TRpS spontaneously.

The crucial point here is that it is, in principle, impossible to specify to which component ${\boldsymbol f}_0^{(i)}$ the theta particle belongs in the $\mu$-space region of overlap between $f^{(1)}_0$ and $f^{(2)}_0$ (see Fig.4).
This point is due to the {\it superposition form} of the distribution ${\boldsymbol f}_0$ attributed to Eq.(\ref{eq:GSdis}), and also due to the singleness of its non-zero fine-grained level $\eta_0$ for the theta particle part (i.e., the singleness of $f_0$).

Due to this point, in the ground state ${\boldsymbol f}_0$, TRp, that is, a replacement of the {\it clock} $t_0$ in the theta particle system---$t_0\to t_0^\prime=F(t_0)$---is required to be applied for both proper times $t^{(i)}$ for $i=1,2$ by a single correspondence rule (i.e., a single time lapse) between clock and proper time due to the singleness of ${\boldsymbol f}_0$.
Thus, one of the two proper times becomes an obvious constraint on TRp for the other proper time.
Due to the discrepancy $N^{(1)}-N^{(2)}$, this constraint breaks TRp invariance of the ground state ${\boldsymbol f}_0$.

\subsection{Decomposition of proper time}

Once TRpS breaks spontaneously, the lapse function in the TRp invariant potential (\ref{eq:pot}) is rearranged to
\begin{equation}
N=N_{\rm g.s.}+\tilde{N}
\end{equation}
for gauge-independent ground state expectation value $N_{\rm g.s.}(>0)$ and fluctuation gauge $\tilde{N}$ with zero temporally local trial mean.
Namely, TRpS is rearranged from a gauge symmetry on gauge $N$ to that on gauge $1+\tilde{N}/N_{\rm g.s.}(>0)$ (note that $\ln N=\ln N_{\rm g.s.}+\ln (1+\tilde{N}/N_{\rm g.s.})$ and $N>0$).\footnote{For the {\it symmetry rearrangement}, see Ref.\cite{Umezawa}.}

Thus, the proper time $t$ is also rearranged to
\begin{equation}
t=t_{\rm g.s.}+\tilde{t}\label{eq:timec}
\end{equation}
for gauge-independent ground state expectation value $t_{\rm g.s.}$ and fluctuation gauge $\tilde{t}$ with zero temporally local trial mean.
This rearrangement of $t$ arises because, according to the ADM spatial coordinate transformation $dx^a\to d\xi^a=dx^a+N^adt_0$ of the square of the line element $ds^2=N^2c^2dt_0^2+q_{ab}d\xi^a d\xi^b$ with the metric signature $(+,-,-,-)$, we set the proper time of the local theta particle system, which is also that of the matter system $\Psi$, as
\begin{equation}
dt=Ndt_0
\end{equation}
by employing an {\it a priori} flowing time $t_0$.
Here, $t_0$ is used in the ADM $3+1$ decomposition of space-time as a reparametrizable time parameter.

In our framework, invoking Sen's picture of gravity as a spin system in $\Sigma_{t_0}$\cite{Sen}, we make the flow of time $t_0$ unique so that this flow is common to all of the clocks $t_0$, $t_0^\prime$, $\ldots$.
This is done instead of making the flow of proper time $t$ unique.
Then, TRp varies the flow of proper time $t$.
Note that if TRpS is not explicitly broken, then each clock with its gauge is equivalent to the other clocks with their respective gauges because TRpS/rearranged TRpS is a gauge symmetry.

Due to this argument and Eq.(\ref{eq:timec}), equivalent increment trials of (rearranged) {\it proper} time, $\delta t$, at an instance fluctuate with a {\it non-zero} variance, $\varsigma$, per the trial mean.
This temporally local fluctuation of proper time flow is attributed to the rearranged TRpS.

Now, in the TRpS unbroken phase, since ground state expectation value of $t$ is a gauge, we must use a gauge-fixed proper time flow, which corresponds to $t_{\rm g.s.}$ in the TRpS broken phase, to describe a time-dependent physical process.
Then, since $\varsigma$ of a gauge-fixed (i.e., unique) proper time flow is zero, the quantum mechanical time evolution of an isolated system in this phase is {\it unitary}.

In the TRpS broken phase, time-dependent physical processes described by using proper time flows $t$ are equivalent to each other with respect to the gauge transformation of proper time
\begin{equation}
t_{\rm g.s.}\to t_{\rm g.s.}+\tilde{t}\;,\label{eq:Gauge}
\end{equation}
which is temporally locally done below the proper time scale $\varsigma\neq 0$.

In the TRpS broken phase, we identify $\varsigma$ with $\sigma$ in Eq.(\ref{eq:SIGMA}).
This identification follows from two facts.
\begin{itemize}
\item Both $\varsigma$, in the TRpS broken phase, and $\sigma$ are the statistical dispersion of equivalent proper time increment trials at an instance and depend on {\it no} Hamiltonian mechanics $\chi(t)$.
Namely, the conversion of time with $\varsigma$ to time with $\sigma$ is not based on a map $t\to \chi(t)$ but is based on the identity map 
\begin{equation}
t\ {\rm with\ temporally\ local\ fluctuations} \to t\ {\rm with\ temporally\ local\ fluctuations}
\end{equation}
in the case of $\varsigma\neq 0$ (i.e., in the TRpS broken phase) and a reduction map 
\begin{equation}
t\ {\rm with\ no\ fluctuations} \to t\ {\rm with\ temporally\ local\ fluctuations}
\end{equation}
in the case of $\varsigma=0$ (i.e., in the TRpS unbroken phase).

\item Both $\varsigma$, in the TRpS broken phase, and $\sigma$ are the upper proper time scale for the physical equivalence of the time-dependent process of the spatial coordinate systems $\{\xi^a_t\}$ under superposition, {\it in the presence of} a group velocity $N_{\rm phys}^a/N$ within space-time $M$.

\end{itemize}

Finally, it is consistent to assume that a certain statistical law governs the behavior of the fluctuation of equivalent proper time increment trials in the TRpS broken phase: there is a finite and definite variance $\sigma$ per the trial mean over all possible instances.
Specifically, we treat $\sigma$ as a physical constant of nature and assume that the proper time increment trials $\delta t$ in this phase follow a normal distribution $\varphi(\delta t)$ with mean $\mu$ and variance $\sigma \mu$.

Since such a statistical law can be set independently from the model in $H_{\rm spin}+H_c$, to test this assumption, we have to rely on experiments, for instance, to measure $t_{\rm life}^{(1)}$, $t_{\rm life}^{(2)}$, $\ldots$ in Eq.(\ref{eq:testex}) for given $(\Delta E)^{(1)}$, $(\Delta E)^{(2)}$, $\ldots$ at instances, respectively.
We can perform such experiments by measuring the time dependence of quantum mechanical interference.

\section{Measurement Processes}

In the Copenhagen interpretation of quantum mechanics, a {\it state reduction} of a pure state consists of two processes: non-selective measurement and the subsequent event reading.
This formulation is in the ensemble interpretation.
In this section, we show that these two processes, with respect to energy measurements, accompany TRpS breaking, which is induced by a matter system $\Psi$.

\begin{figure}[htbp]
\begin{center}
\includegraphics[width=0.32\hsize,bb=4 4 255 255]{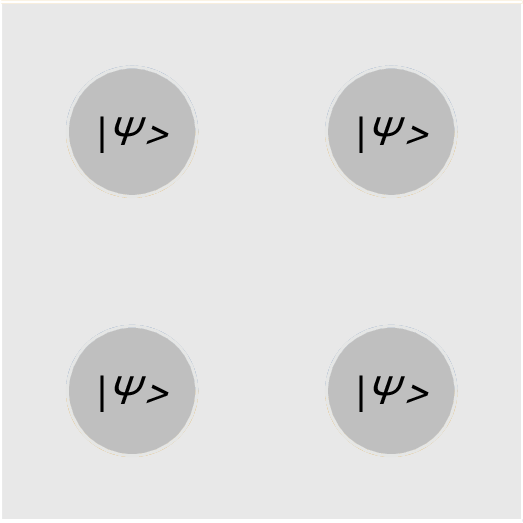}
\includegraphics[width=0.32\hsize,bb=4 4 255 255]{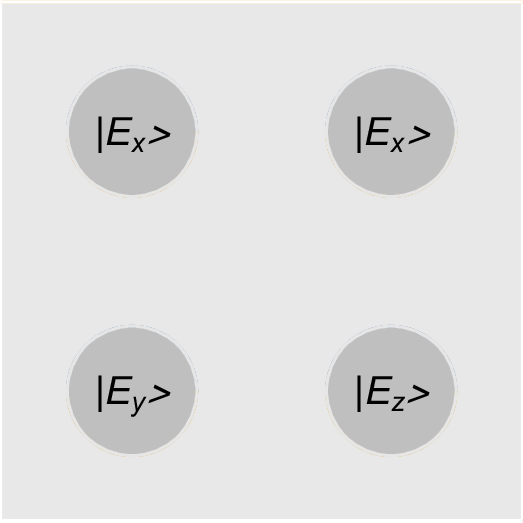}
\includegraphics[width=0.32\hsize,bb=4 4 255 255]{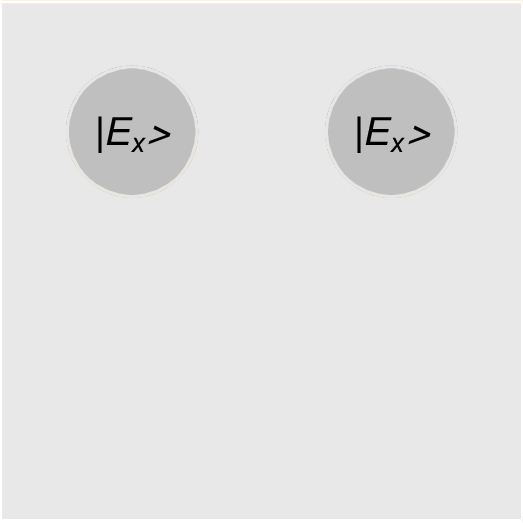}
\end{center}
\caption{Schematics of the changes of the ensemble of four copies of the matter system $\Psi$ during the energy measurement process in the Copenhagen interpretation.
The left panel shows the quantum pure ensemble of $\Psi$ with the initial state vector $|\Psi\ra=|E_x\ra/\sqrt{2}+|E_y\ra/2+|E_z\ra/2$ for three energy eigenstates $|E_x\ra$, $|E_y\ra$, and $|E_z\ra$ with distinct energy eigenvalues.
The middle panel shows the classical mixed ensemble of $\Psi$ after the {\it non-selective energy measurement}.
The right panel shows the classical pure ensemble of $\Psi$ after an {\it energy event reading} with the measurement event $E_x$.
The probability of obtaining this event is $1/2$.}
\end{figure}

In Fig.5, we show schematically these two processes, with respect to energy measurements, applied to the quantum pure ensemble of an initial pure state $|\Psi\ra$ of the matter system $\Psi$.

In a broad sense, this state reduction is the part (of a quantum measurement) for the {\it event reading system} $\Psi$.
To complete the description of a quantum measurement, we need also to consider the part for the {\it measured system} and the measurement apparatus.
For a complete description of direct quantum measurements (i.e., error-free/projective quantum measurements), see Appendix B.

\subsection{Non-selective energy measurement: the middle panel in Fig.5}

In time-dependent processes of the matter system $\Psi$ with its Hamiltonian $\widehat{{{H}}}$, the {\it non-zero} statistical variance per the trial mean of proper time increment trials that accompanies TRpS breaking is represented by the non-selective energy measurement of the state vector of the system $\Psi$ in the Schr$\ddot{{\rm{o}}}$dinger picture, as seen in the following way.

In the Schr$\ddot{{\rm{o}}}$dinger picture, the time-evolution equation of the state vector $|\Psi(t)\ra$ of the matter system $\Psi$ is the Schr$\ddot{{\rm{o}}}$dinger equation.
We write it in the time increment formalism using the proper time of the matter system $\Psi$ with broken TRpS as
\begin{equation}
i\hbar\frac{\delta |\Psi(t)\ra}{\delta t}=\widehat{{{H}}}|\Psi(t)\ra\;.\label{eq:Sch}
\end{equation}
In the history representation, its solution can be written as
\begin{eqnarray}
|\Psi(t)\ra=\sum_nc_ne^{-\frac{it}{\hbar}E_n}|E_n\ra\label{eq:PsiE}
\end{eqnarray}
in the energy eigenbasis $\{|E\ra\}$.

For this state vector, the average of the expectation value of an arbitrarily given observable $\widehat{A}$ of the matter system $\Psi$ with respect to proper time increment trials $\delta t$ is
\begin{eqnarray}
\overline{\la \Psi(\mu)|\widehat{A}|\Psi(\mu)\ra}&=&\sum_{m,n} c_m\bar{c}_n\int d (\delta t^\prime) \varphi(\delta t^\prime)e^{-\frac{i\delta t^\prime}{\hbar}(E_m-E_n)}{\la E_n|\widehat{A}|E_m\ra}\label{eq:PSI2}\\
&=&\sum_{m,n}c_m\bar{c}_n e^{-\frac{i\mu}{\hbar}(E_m-E_n)}e^{-\frac{\sigma \mu}{2\hbar^2}(E_m-E_n)^2}{\la E_n|\widehat{A}|E_m\ra}\label{eq:sigma2}
\end{eqnarray}
when the proper time increment trials $\delta t$ follow the normal distribution $\varphi(\delta t)$ with mean $\mu$ and variance $\sigma \mu$\cite{4model}.
Here, the distribution $\varphi(\delta t)$ was introduced in section 3.6.

This formula gives the time evolution of the averaged expectation value of the observable $\widehat{A}$ in the history representation
\begin{equation}
\overline{\la \Psi(t)|\widehat{A}|\Psi(t)\ra}=\sum_{m,n}c_m\bar{c}_ne^{-\frac{it}{\hbar}{(E_m-E_n)}}e^{-\frac{\sigma t}{2\hbar^2}{(E_m-E_n)}^2}{\la E_n|\widehat{A}|E_m\ra}\;,\label{eq:sol}
\end{equation}
where the exponential damping factors for $m$ and $n$ (s.t. $E_m\neq E_n$) give rise to the non-selective energy measurement (i.e., quantum energy coherence of $|\Psi(t)\ra$ with respect to $\{\overline{\la\widehat{A}\ra}\}$ becomes unobservable).

Here, the lifetime, $t_{\rm life}$, of quantum energy coherence of $|\Psi(t)\ra$ with respect to $\{\overline{\la\widehat{A}\ra}\}$ is equal to the characteristic time of exponentially decaying off-diagonal elements of the averaged density matrix $\overline{|\Psi(t)\ra\la\Psi(t)|}$ expressed in the energy eigenbasis $\{|E\ra\}$.
From this, we obtain
\begin{equation}
t_{\rm life}=\frac{2\hbar^2}{\sigma (\Delta E)^2}\label{eq:testex}
\end{equation}
for the energy difference $\Delta E$ in each off-diagonal element\cite{4model}.

\subsection{Energy event reading: the right panel in Fig.5}

Now, we consider a state change with no dynamical element, which is attributed to the equivalence with respect to the gauge transformation of proper time (\ref{eq:Gauge}), as an informatical process.
For two different proper time increment trials $\delta t_1$ and $\delta t_2$, the equivalence $\simeq$ of two state vectors up to an overall phase-factor difference
\begin{equation}
e^{-\frac{i\delta t_1}{\hbar}\widehat{H}}|\Psi(0)\ra\simeq e^{-\frac{i\delta t_2}{\hbar}\widehat{H}}|\Psi(0)\ra\;,\ \ \delta t_1\neq \delta t_2\label{eq:PSI}
\end{equation}
holds if and only if the state vector $|\Psi(0)\ra$ is an {\it energy eigenstate}.
Indeed, for an energy superposition
\begin{equation}
|\Psi(0)\ra=c_1|E_1\ra+c_2|E_2\ra\;,\ \ E_1\neq E_2\;,
\end{equation}
there is discrepancy ratio $e^{-i(\delta t_1-\delta t_2)(E_1-E_2)/\hbar}$ between the relative phases in the left- and right-hand sides of Eq.(\ref{eq:PSI}):
\begin{eqnarray}
e^{-\frac{i\delta t_1}{\hbar}\widehat{H}}|\Psi(0)\ra&=&c_1e^{-\frac{i \delta t_1}{\hbar}E_1}|E_1\ra+c_2e^{-\frac{i \delta t_1}{\hbar}E_2}|E_2\ra\;,\\
e^{-\frac{i\delta t_2}{\hbar}\widehat{H}}|\Psi(0)\ra&=&c_1e^{-\frac{i \delta t_2}{\hbar}E_1}|E_1\ra+c_2e^{-\frac{i \delta t_2}{\hbar}E_2}|E_2\ra\;.
\end{eqnarray}

When there exists quantum energy coherence of $|\Psi(0)\ra$ with respect to the averaged expectation values of all observables $\{\overline{\la \widehat{A}\ra}\}$ of the matter system $\Psi$, the fluctuation of equivalent proper time increment trials cannot be described by using $|\Psi(0)\ra$ or imposing Eq.(\ref{eq:PSI}) on $|\Psi(0)\ra$ (instead, $\{\la \widehat{A}\ra\}$ are replaced with $\{\overline{\la \widehat{A}\ra}\}$).
This is because Eq.(\ref{eq:PSI}) is the condition for a state change with no dynamical element (i.e., no change of time) and thus it cannot extinguish the quantum energy coherence.

However, just after a non-selective energy measurement of a pure state $|\Psi\ra$ (e.g., the time evolution resulting from Eq.(\ref{eq:sol})), viewed as the preparation of the state $|\Psi(-0)\ra$ just before the informatical process induced by Eq.(\ref{eq:PSI}) at $t=0$, quantum energy coherence of $|\Psi(-0)\ra$ with respect to $\{\overline{\la\widehat{A}\ra}\}$ cannot be observed (i.e., for $\widehat{A}$, its total interference is less than its uncertainty width $\Delta \widehat{A}$, with respect to the decohered state $|\Psi(-0)\ra$\cite{Konishi2}):
\begin{equation}
\bigl(\Delta \widehat{A}\bigr)^2\gg \Biggl(2\sum_{m<n}{\rm Re}\bigl[\overline{\la \Psi_n|\widehat{A}|\Psi_m\ra}\bigr]\Biggr)^2\ \ {\rm for}\ {\rm all}\ \ \widehat{A}\;,
\end{equation}
where $\{|\Psi_n\ra\}$ are the vector components of the state vector $|\Psi(-0)\ra$ in the energy eigenbasis.
Namely, this state $|\Psi(-0)\ra$ is {\it equivalent to} a classical mixed ensemble of energy eigenstates in the ensemble interpretation with respect to $\{\overline{\la\widehat{A}\ra}\}$.

At such a time, for the {\it state vector} $|\Psi(0)\ra$, Eq.(\ref{eq:PSI}) induces the informatical process, that is, a quantum state reduction that satisfies the Born rule as a result of the preceding non-selective energy measurement: this is the subsequent energy event reading (i.e., elimination of the other energy events from the classical mixed ensemble).

\section{Summary and Discussion}

In the Copenhagen interpretation of quantum mechanics, a state reduction of a pure state consists of two processes: non-selective measurement and the subsequent event reading.
This formulation is in the ensemble interpretation.
In this paper, we proposed, for the first time, a mechanism for event reading, with respect to energy.
In this mechanism, energy event reading occurs in a macroscopic {\it coherence domain}, in which a spatially macroscopically coherent quantum matter system exists, due to the fluctuation of equivalent proper time increment trials attributed to system-induced spontaneous breaking of TRpS in canonical gravity.

This mechanism works in the present framework of an extended model of canonical gravity theory, under two assumptions:
\begin{enumerate}
\item[A1] A novel interpretation of quantum mechanics with respect to the quantum mechanical position uncertainties of particles, where we reverse the roles of particles and the spatial configuration space in the uncertainties of the positions of particles, and the wave function of particles in this novel interpretation is the same as that in the conventional interpretation.

\item[A2] The hypothesis of the existence of one additional massive spin-$1/2$ particle (we call it the {\it theta particle}) as CDM having a long-range spin-spin self-interaction and a spin-spin interaction (mathematically, in the form of an exchange interaction) with the spinorial expression of the spatial-diffeomorphism-gauge independent part of the shift vector, which is attributed to the assumption A1.
\end{enumerate}

Here, we summarize the contents of the proposed scheme of energy measurement processes of a desired quantum matter system $\Psi$ briefly.
This is accomplished through the following three steps.

In the first step, the {\it superposed form} collisionless equilibrium (i.e., double Lynden-Bell) distribution (\ref{eq:DLB}) of a local long-range self-interacting theta particle system is dynamically generated by the Hamiltonian $H_{\rm spin}$ of the theta particle field from the initial magnetized distribution of the theta particles.
This initial distribution of the theta particles is prepared by the matter system $\Psi$, provided that $\Psi$ is weakly coupled and spatially macroscopically coherent.

This preparation of the initial distribution of the theta particles is the result of
(i) interactions between each theta particle and its environmental particles and
(ii) the ferromagnetic coupling $H_c$ of the spin vector of each theta particle with the overall aligned configuration of the spatial-diffeomorphism-gauge independent part of the shift vector $N^{(\alpha\beta)}_{\rm phys}=c\lambda^{\dagger(\alpha}\lambda^{\beta)}$ within the coherence domain of the matter system $\Psi$.

The overall aligned configuration of $N^a_{\rm phys}$ is attributed to the two provided properties of $\Psi$ and our novel interpretation of quantum mechanics A1.

In the second step, the generated local theta particle system converts the {\it a priori} flowing time $t_0$ into a proper time $t$ with increment trials, which fluctuate with a non-zero variance per the trial mean, by the spontaneous breaking of TRpS.
This spontaneous TRpS breaking is due to the existence of the TRp invariant double Lynden-Bell potential (\ref{eq:pot}) of the total spin system, where the time lapse is rescaled in the {\it superposed form} ground state (\ref{eq:Gstate}) in two distinct ways (see Eq.(\ref{eq:ti})).

[On the other hand, spatial diffeomorphism invariance is not broken.
The reason why spatial diffeomorphisms behave differently from TRp is that the part $H_c$ of the total Hamiltonian of the whole system is the TRp invariant potential but is spatial-diffeomorphism-gauge independent.]

In the third step, by solving the Schr$\ddot{{\rm{o}}}$dinger equation (\ref{eq:Sch}) of the state vector (written in the time increment formalism), we have shown that this conversion of time (\ref{eq:timec}) results in non-selective measurements and event readings of the matter system $\Psi$ with respect to its energy.

In this third step, whereas non-selective energy measurement is a known result\cite{4model}, energy {\it event reading} is a novel result.

In the following, we discuss our model from three aspects: its advantage, its experimental tests, and the deducible properties of the additional particles.

First, we clarify the advantage of our model of extended canonical gravity in comparison with canonical gravity.
In canonical gravity, selection of time is blocked by TRpS, which is a gauge symmetry of the totally constrained system.
In our model, on the other hand, this unsatisfactory situation is resolved by breaking TRpS spontaneously.
This resolution does not suppose quantum measurements {\it a priori}.
Notably, in our model, the quantum measurements in the Copenhagen interpretation are the consequence of TRpS breaking.

Second, our extension of canonical gravity alters general relativity not in the gravitational force itself but in the space-time diffeomorphism invariance related to the concept of time.
In principle, this extension of canonical gravity can be experimentally tested through its three windows.
The first window is the observation of theta particles and anti-theta particles.
The second window is the non-selective energy measurement, in which the quantum mechanical interference decays as a dynamical process.
The third window is the partial and explicit breaking of the space-time diffeomorphism invariance, which was explained in section 2.3.3.
Of course, the first two prior experimental tests, namely, the realization of the Copenhagen interpretation of quantum mechanics and the selection of time, are excluded from this list.

Third, in the present model, we can deduce the qualitative properties of the theta particle when it is interpreted as CDM.

(i) We assume that the theta particle is electrically neutral.
Then, it would not radiate photons.
In the present Dirac fermion model, we assume that the theta particle has at least one kind of gauge interaction so that it is not truly neutral.

(ii) The theta particle is a non-relativistic massive particle.
The local density of CDM is observed to be on the order of 0.1 GeV$\cdot$cm$^{-3}$\cite{Bertone}.
In order to have a high-number-density profile, the theta particle needs to be light CDM like the axion.
Here, the axion is the ultra-light CDM candidate with mass on the order of $\mu$eV to meV\cite{Feng} and the mean particle number per 1 cm$^3$ on the order of $10^{11}$ to $10^{14}$.

Finally, it is worthwhile to discuss the novelty of our theory of measurement by comparing it with our prototype, the von Neumann--Wigner (vNW) theory of measurement\cite{Neumann,Wigner}.

One can raise two objections to the vNW theory.
First, the {\it projection hypothesis} that plays the central role in this theory is unphysical and was regarded as the {\it `abstract ego' of an observer}\cite{Neumann}.
Second, even if one admits the {\it projection hypothesis}, the vNW theory does not include a non-selective measurement process as a step.
In regard to the latter fact, the state reduction in the vNW theory is not properly decomposed into dynamical and informatical parts.
This means that the vNW theory is not compatible with the ensemble interpretation.
Of course, the ensemble interpretation does not rule out the Copenhagen interpretation.

To address these two shortcomings of the vNW theory, first, the {\it projection hypothesis} in our theory of measurement is physical and is realized as the {\it theta particle system confined in a coherence domain}.
Second, in our mechanism, state reduction is properly decomposed into a dynamical part (the non-selective measurement) and an informatical part (the event reading).
Our theory is thus compatible with the ensemble interpretation.

\begin{appendix}

\section{Time Concept in Canonical Gravity}

In this Appendix, we give a simple overview of the time concept in canonical gravity {\it before its extension}.

In this paper, to study the time evolution of space-time structure, we use the ADM $3+1$ decomposition of a space-time $(M,g_{\mu\nu})$\cite{ADM}.
In the ADM $3+1$ decomposition method, the square of the line element in $M$ is written as
\begin{equation}
ds^2=N^2c^2dt_0^2+q_{ab}(dx^a+N^a dt_0)(dx^b+N^bdt_0)\label{eq:ADMapp}
\end{equation}
for time-lapse function $N$ and shift vector $N^a$.

Using Eq.(\ref{eq:ADMapp}), we obtain the gravitational Hamiltonian of a space-like hypersurface $\Sigma_{t_0}$ from the Einstein-Hilbert action as
\begin{equation}
H_{\rm grav}=\int_{\Sigma_{t_0}}(Nh_{{\rm grav}}-N_ah^a_{\rm grav})dS\;,\label{eq:HADM}
\end{equation}
where
\begin{eqnarray}
h_{\rm grav}&=&\sqrt{-q}(-{}^{(3)}R-K^{ab}K_{ab}+K^2)\;,\label{eq:SUPERH}\\
h^a_{\rm grav}&=&2\sqrt{-q}D_b(K^{ab}-Kq^{ab})\label{eq:SUPERM}
\end{eqnarray}
for scalar curvature ${}^{(3)}R$ and covariant derivative $D_a$ on $(\Sigma_{t_0},q_{ab})$ and $K=K^{ab}q_{ab}$.

Both a {\it super-Hamiltonian} $h_{\rm grav}=h_{{\rm grav},t_0}(x)$ and a {\it supermomentum} $h^a_{\rm grav}=h^a_{{\rm grav},t_0}(x)$ are set to vanish by including the contributions from the other sectors as the super-Hamiltonian and supermomentum constraints with Lagrange multipliers $N$ and $N^a$, respectively; otherwise, the space-time diffeomorphism invariance of the total Hamiltonian is explicitly broken.
In canonical gravity, these four constraints automatically hold at all times if all of these hold at the initial time\cite{Geometry}.

Intuitive interpretations of Eqs.(\ref{eq:ADMapp}) and (\ref{eq:HADM}) are as follows.

First, the phase-space variables of canonical gravity are the induced metric tensor $q_{ab}$ and its canonical conjugate.

Second, regarding $\Sigma_{t_0}$ as the embedding of a space manifold $S$ into space-time $M$, denoted by $e=e_{t_0}(x)$, the lapse function $N$ and the shift vector $N^a$ are the components of the deformation of the embedding $\pa e/\pa t_0$ perpendicular and parallel to $\Sigma_{t_0}$, respectively\cite{Kuchar}.

Third, the lapse function $N$ and the shift vector $N^a$ do not appear in the super-Hamiltonian (\ref{eq:SUPERH}) or the supermomentum (\ref{eq:SUPERM}) and are the time-dependent temporal gauge variables of time reparametrization and the space diffeomorphism of $S$, respectively.

Fourth, the arbitrariness of the four variables $N$ and $N^a$ in the space-time metric tensor $g_{\mu\nu}$ reflects the arbitrariness of choice of space-time coordinate system.

\section{Model of Direct Quantum Measurements}

In this Appendix, we give a model of direct quantum measurements following Refs.\cite{Konishi2,Konishi}.\footnote{The contents of this Appendix are not new and accord with Refs.\cite{Konishi2,Konishi}, and there is no link between the contents of this Appendix and the extension of canonical gravity.}
Because energy does not have a canonical conjugate, error-free entanglement between
(i) a given quantum measured system $\psi$ with eigenstates of measured discrete observable $\widehat{{\cal O}}$ and
(ii) the event reading system $\Psi$ with energy eigenstates
is taken by a non-selective ${\cal O}$-measurement-based quantum energy feedback in the open quantum system $\psi+A+\Psi$, where $A$ is a {\it macroscopic} measurement apparatus.

Here, a non-selective ${\cal O}$-measurement can be realized by a von Neumann-type interaction $\widehat{H}_{\rm int}^{\psi A}$ (see Eq.(\ref{eq:vNint})) between the measured system $\psi$ and the measurement apparatus $A$ in a unitary process $\widehat{U}^{\psi A}$ that is quantum mechanically long-term but macroscopically instantaneous, with the state change of $A$ unobservable\cite{Araki,Konishi2,Konishi,Ozawa}.

In this model, $A$ is characterized such that it has one degree of freedom, with a center of mass position that cannot be sharply measured and a center of mass momentum, $\widehat{P}$, assumed to be the continuous superselection rule observable\cite{Araki}.
Here, a {\it continuous superselection rule observable} represents a {\it classical observable}\cite{Araki} and is characterized by two properties: having a virtually continuous spectrum (i.e., being able to be very sharply but not exactly measured) and commuting with (i.e., being able to be simultaneously measured, {\it within certain measurement errors}, with) all orbital observables, $\{\widehat{O}\}$, of $A$ which are selected by this superselection rule $[\widehat{O},\widehat{P}]=0$\cite{Neumann,Araki}.
Note that, within certain measurement errors, all orbital observables of $A$ commute with each other, and the full set of simultaneous eigenfunctions of the redefined center of mass position and momentum operators forms a complete orthonormal system of the functional Hilbert space of $A$: this is von Neumann's theorem in Ref.\cite{Neumann}.

For this property of $A$, a generic pure state $|\psi\ra|A_0\ra$ of the combined system $\psi+A$ can be well described by a direct integral,
\begin{equation}
|\psi\ra|A_0\ra=\int^\bigoplus|\psi(p)\ra A_0(p)dp\;,\label{eq:direct}
\end{equation}
where a state vector $|\psi(p)\ra\equiv |\psi\ra|p\ra$ belongs to the continuous superselection sector specified by an eigenvalue $p$ of the continuous superselection rule observable $\widehat{P}$, and $A_0(p)$ is the wave function of $|A_0\ra$ in the $p$-representation\cite{Araki}.
Here, for two state vectors belonging to different continuous superselection sectors, the corresponding matrix elements of {\it any} observable of the combined system $\psi+A$ are all identically zero\cite{Araki}: there is no quantum coherence between these two state vectors.
This is because $\widehat{P}$ commutes with any orbital observable $\widehat{O}$ of $A$.

Under this structure (\ref{eq:direct}) of a given pure state $|\psi_0\ra|A_0\ra$ of the combined system $\psi+A$, the expectation value of an arbitrarily given observable of this system, selected by the superselection rule,
\begin{equation}
\widehat{{\cal X}}=\int^\bigoplus \widehat{{\cal X}}(p)dp
\end{equation}
is a quantity averaged over $p$ using the weight $|A_0(p)|^2$\cite{Araki}, and a non-selective ${\cal O}$-measurement with respect to $\{\la \widehat{{\cal X}}\ra\}$ is realized as a quantum mechanically long-term but macroscopically instantaneous dynamical process by a von Neumann-type interaction\cite{Neumann,Konishi2},
\begin{eqnarray}
\widehat{H}_{\rm int}^{\psi A}&=&-\Lambda \widehat{{\cal O}}\otimes \widehat{P}\label{eq:vNint}\\
&=&-\Lambda \int^\bigoplus \widehat{{\cal O}}(p)p dp\;,\ \ \widehat{{\cal O}}(p)\equiv \widehat{{\cal O}}\otimes |p\ra\la p|\;,
\end{eqnarray}
in which $\Lambda$ is strong enough that we can neglect the kinetic Hamiltonians of the systems $\psi$ and $A$\cite{Ozawa}.\footnote{In this argument, the Riemann-Lebesgue theorem is used (see Appendix D in Ref.\cite{Konishi}).}

Now, the whole system of processes is
\begin{eqnarray}
|\psi_0\ra|A_0\ra|E_0\ra&=&\Biggl(\sum_n c_n|{\cal O}_n\ra\Biggr)|A_0\ra|E_0\ra\ \ ({\rm initial\ quantum\ pure\ state})\label{eq:prepreFB}\\
&\to&\sum_n c_n \widehat{U}^{\psi A}|{\cal O}_n\ra|A_0\ra|E_0\ra\ \ ({\rm{non\ selective\ {\cal O}\ measurement}})\label{eq:preFB}\\
&\to&\sum_nc_n \widehat{U}^{\psi A}|{\cal O}_n\ra|A_0\ra|E_n\ra\ \ ({\rm quantum\ energy\ feedback})\;.\label{eq:FB}
\end{eqnarray}

(Note that whereas the state in Eq.(\ref{eq:prepreFB}) is a quantum pure state, the states in Eqs.(\ref{eq:preFB}) and (\ref{eq:FB}) under the structure (\ref{eq:direct}) are {\it equivalent to} classical mixed ensembles of product ${\cal O}$-eigenstates in the ensemble interpretation with respect to the expectation values of all observables of the combined system $\psi+A+\Psi$.)

In Eq.(\ref{eq:FB}), the quantum energy feedback process in the open quantum system $\psi+A+\Psi$ is controlled by a feedback controller; before and after the quantum energy feedback process, the feedback controller and the energy reservoir are {\it decoupled} from (i.e., in a product state with) the combined system $\psi+A+\Psi$ because this process is subsequent to the non-selective ${\cal O}$-measurement in the combined system $\psi+A$\cite{Konishi2}.
Then, the feedback controller and the energy reservoir can be traced out without changing the state of the combined system $\psi+A+\Psi$\cite{Konishi2}.

After this entanglement (\ref{eq:FB}), the event reading in $\Psi$ with respect to energy agrees with the event reading in $\psi+A+\Psi$ with respect to ${\cal O}$ and energy.

\end{appendix}

\end{document}